\documentstyle[prb,aps,epsfig,eqsecnum,amsmath,amssymb]{revtex}%%prp%%
\begin{document}
%-------------------------------------------------------------------------------
%one-column-prb-style for the title and abstract--> goc-to-prb or prb-to-goc
\twocolumn[\hsize\textwidth\columnwidth\hsize\csname@twocolumnfalse%%prb%%
\endcsname%%prb%%
%-------------------------------------------------------------------------------
\title{Soft modes, Gaps and Magnetization Plateaus in 1D Spin-1/2
  Antiferromagnetic Heisenberg Models} 
\author{A. Fledderjohann, C. Gerhardt, M. Karbach, K.-H. M\"utter 
        and R. Wie\ss ner}
\address{Physics Department, University of Wuppertal, 42097 Wuppertal, Germany} 
\date{\today-v2.2}
\maketitle
%\twocolumn
%%%%%%%%%%%%%%%%%%%%%%%%%%%%%%%%%%%%%%%%%%%%%%%%%%%%%%%%%%%%%%%%%%%%%%%%%%%%%%%%
%
\begin{abstract}
%
%%%%%%%%%%%%%%%%%%%%%%%%%%%%%%%%%%%%%%%%%%%%%%%%%%%%%%%%%%%%%%%%%%%%%%%%%%%%%%%%
  We study the one-dimensional spin-1/2 model with nearest and
  next-to-nearest-neighbor couplings exposed to a homogeneous magnetic
  field $h_{3}$ and a dimer field with period $q$ and strength $\delta$. The
  latter generates a magnetization plateau at $M=(1-q/\pi)/2$, which evolves
  with strength $\delta$ of the perturbation as $\delta^{\epsilon}$, where
  $\epsilon=\epsilon(h_{3},\alpha)$ is related to the $\eta$-exponent which
  describes the critical behavior of the dimer structure factor, if the
  perturbation is switched of ($\delta=0$). We also discuss the appearance of
  magnetization plateaus in ladder systems with $l$ legs.
\end{abstract}
%\pagebreak
%insert suggested PACS number in the next line
\pacs{??}
%\pagebreak
\twocolumn%%prb%%
%-------------------------------------------------------------------------------
%end of one column-prb-style
]%%prb%%
%-------------------------------------------------------------------------------
%gerhard-one-column-draft style, centered
%%goc%%\newpage
%%goc%%\narrowtext
%%goc%%\oddsidemargin=4cm
%%prp%%\onecolumn
%%%%%%%%%%%%%%%%%%%%%%%%%%%%%%%%%%%%%%%%%%%%%%%%%%%%%%%%%%%%%%%%%%%%%%%%%%%%%%%%
%
\section{Introduction}
%
%%%%%%%%%%%%%%%%%%%%%%%%%%%%%%%%%%%%%%%%%%%%%%%%%%%%%%%%%%%%%%%%%%%%%%%%%%%%%%%%
The existence or nonexistence of a gap between the energies of the ground state
and the low lying excited states is the most important criterium for the
criticality of a quantum spin system. Haldane's conjecture \cite{Hald83,Hald83a}
states that one-dimensional (1D) quantum spin systems have no gap for half
integer spin $s$, but do have a gap for integer spin $s$. The
conjecture only holds for an appropriate choice of the couplings of the spins at
nearest neighbor sites.\cite{Muet94}

The Lieb, Schultz, Mattis (LSM) construction\cite{LSM61,AL86} allows rigorous
statements on the degeneracy of the ground state. Starting from the unitary
operator
\begin{equation}
 {\bf U} \equiv\exp \left(-i\frac{2\pi}{N} \sum_{l=1}^{N} l S^{3}_{l}\right),
\label{eq:i1}
\end{equation}
it is straightforward to prove that the application of the operator ${\bf
  U}^k,\; k=1,2,... $ on the ground state $|0\rangle$ generates {\it new} states
\begin{equation}
 |k\rangle \equiv {\bf U}^k |0\rangle \quad k=1,2,\ldots;\; k \mbox{ finite},
\label{eq:i2}
\end{equation}
with an energy expectation value
\begin{equation}
\langle  k|{\bf H}|k \rangle - \langle 0|{\bf H}|0\rangle = O(N^{-1})
\label{eq:i3}
\end{equation}
approaching the ground state energy $E_0 \equiv \langle 0|{\bf H}|0\rangle$ in the
thermodynamical limit $N \to \infty$.

Of course the crucial question is whether the {\it new} states $|k\rangle $ are
different from the ground state $|0\rangle$ or not. This question can be
answered by an analysis of the quantum numbers of the states $|k\rangle,\;
k=1,2,\ldots$. For example in the case of the Spin-1/2 Hamiltonian with nearest
neighbor couplings:
\begin{equation}
   {\bf H}(h_{3}) \equiv 2\sum_{l=1}^{N} {\bf S}_{l} \cdot {\bf S}_{l+1}  
     -2h_{3} {\bf S}_3(0),
\label{eq:i4}
\end{equation}
and
\begin{equation}
{\bf S}_a(q) \equiv \sum_{l=1}^{N} e^{iql} S_l^{a},\quad a=1,2,3,
\label{eq:i5}
\end{equation}
the ground state has momentum $p_{s}=0,\pi$ and the total Spin $S_{T}^{3}=S=N
M(h_{3})$, where $M$ is the magnetization.  The new states $|k\rangle$ turn out
to be eigenstates of the translation operator ${\bf T}$:
\begin{equation}
{\bf T} |k\rangle = {\bf T} {\bf U}^k |0\rangle = e^{i p_k} |k\rangle,
\label{eq:i6}
\end{equation}
where \cite{OYA97}
\begin{equation}
  p_k = p_{s} + k q_3(M),
\label{eq:i7}
\end{equation}
and $q_3(M)\!\equiv\!\pi (1-2M)$.  For $M=0$ the ground state $|0\rangle$ and
the new state $|1\rangle$ differ in their momenta by $\pi$ and are therefore
orthogonal to each other.  \\
For $M=1/4$ one finds a fourfold degeneracy of the ground state with momenta
$p_k=p_{s} + k \pi/2,\; k=0,1,2,3 $.

It should be noted that the LSM construction allows to identify the zero
frequency excitations (soft modes) in the model with Hamiltonian (\ref{eq:i4}).
Some of these soft modes induce characteristic signatures, e.g. zeroes in the
dispersion curve and singularities in the transverse and longitudinal structure
factors at the soft mode momenta $q=q_1(M)\equiv\pi,\; q=q_3(M)$, which can be
easily recognized even on rather small systems.\cite{KMS95} Following conformal
field theory the corresponding critical $\eta$-exponents can be determined from
the finite-size behavior of the dispersion curve at the soft mode
momenta.\cite{FGM+96} It is known that the $\eta$-exponents of the soft mode --
i.e. the $M$ dependence of $\eta_3(M),\eta_1(M)$ -- changes, \cite{GFA+97} if
we add further couplings to the Hamiltonian (\ref{eq:i4}). In some cases (see
the discussion below) the soft mode might disappear completely and a gap opens
between the states, which were gapless before switching on the perturbation.

The following cases (A.-E.) have been studied so far.
%%%%%%%%%%%%%%%%%%%%%%%%%%%%%%%%%%%%%%%%%%%%%%%%%%%%%%%%%%%%%%%%%%%%%%%%%%%%%%%%
\subsection{A transverse staggered field}\label{sec:S1q}
%%%%%%%%%%%%%%%%%%%%%%%%%%%%%%%%%%%%%%%%%%%%%%%%%%%%%%%%%%%%%%%%%%%%%%%%%%%%%%%%
A gap was found \cite{FKM98a,FKM98b,OA97} in a transverse staggered field of
strength $h_{1} {\bf S}_1(\pi)$,
\begin{equation}
  \label{eq:Hh3h1}
     {\bf H}(h_{3},h_{1}) \equiv {\bf H}(h_{3})+2h_{1} {\bf S}_1(\pi),
\end{equation}
between the states which differ in their momenta by $\pi$. Indeed the operator
${\bf S}_1(\pi)$ is invariant only under translations ${\bf T}^2$ and the
eigenstates of the Hamiltonian are only eigenstates of ${\bf T}^2$.
  
In the free field case ($h_{3}=0$) the ${\bf T}^2$ quantum numbers of the ground
state $|0\rangle$ and of the LSM state $|1\rangle={\bf U} |0\rangle$ are the
same and the twofold degeneracy of the ground state is lifted by the explicit
breaking of translation invariance.
  
The fourfold degeneracy with momenta $p\!=\!0,\pi,\pm\!\pi/2$ which occurs at
$M=1/4$ and $h_{1}=0$ is lifted in the following manner. The states with $p=0$
and $p=\pi$ are even with respect to ${\bf T}^{2}$. The same holds for the
ground state $|0\rangle$, which is a linear combination of $p=0$ and $p=\pi$
components. A gap opens to the second state, which is even under ${\bf T}^{2}$.
The gap evolves with the strength $h_{1}$ of the perturbation as
$h_{1}^{\epsilon}$. The exponent $\epsilon=\epsilon_{1}(h_{3})$ is
given by the exponent $\eta_1(M)$
\begin{equation}
  \epsilon_1(h_{3}) = 2[4-\eta_1(M(h_{3}))]^{-1},
  \label{eq:i9}
\end{equation}
associated with the divergence of the transverse structure factor at $q=\pi$.
  
The LSM construction with the operator (\ref{eq:i1}) leads to a second state
$|1\rangle = {\bf U} |0 \rangle$ which is degenerate with the ground state
$|0\rangle$ and which is odd under ${\bf T}^2$. This state can be constructed as
a linear combination of momentum eigenstates with $p=\pm\pi/2$.
%%%%%%%%%%%%%%%%%%%%%%%%%%%%%%%%%%%%%%%%%%%%%%%%%%%%%%%%%%%%%%%%%%%%%%%%%%%%%%%%  
\subsection{A longitudinal periodic field}\label{sec:bS3q}
%%%%%%%%%%%%%%%%%%%%%%%%%%%%%%%%%%%%%%%%%%%%%%%%%%%%%%%%%%%%%%%%%%%%%%%%%%%%%%%%
A longitudinal periodic field ${\bf \bar S}_3(q)$ of strength $2h_{q}$
\begin{equation}
  \label{eq:100}
     {\bf H}(h_{3},h_{q}) \equiv {\bf H}(h_{3})+2h_{q} {\bf \bar S}_3(q),
\end{equation}
with
\begin{equation}
  {\bf \bar S}_3(q)\equiv[{\bf S}_3(q) + {\bf S}_3(-q)]/2,
  \label{eq:i10}
\end{equation}
induces a plateau in the magnetization curve $M=M(h_{3})$ at
$M =(1-q/\pi)/2$, i.e. $q$ has to meet the soft mode momentum $q=q_3(M)$. The
difference of the upper and lower critical field:
\begin{equation}
  \Delta({h_{q}},h_{3})\equiv 
  h^{u}_{3} - h^{l}_{3} \sim h_{q}^{\epsilon_3(h_{3})},
\end{equation}
evolves with an exponent, which is again related via (\ref{eq:i9}) to the
corresponding $\eta_3$-exponent and which can be extracted from the finite-size
behavior of the longitudinal structure factor \cite{FGM+96} at $q=q_3(M)$.
%%%%%%%%%%%%%%%%%%%%%%%%%%%%%%%%%%%%%%%%%%%%%%%%%%%%%%%%%%%%%%%%%%%%%%%%%%%%%%%%  
\subsection{A next-to-nearest-neighbor coupling}\label{sec:Oalpha}
%%%%%%%%%%%%%%%%%%%%%%%%%%%%%%%%%%%%%%%%%%%%%%%%%%%%%%%%%%%%%%%%%%%%%%%%%%%%%%%%
A next-to-nearest-neighbor coupling
\begin{equation}
  {\bf H}_{2} \equiv 2\sum_{l=1}^{N} {\bf S}_{l} \cdot{\bf S}_{l+2},
  \label{eq:i12}
\end{equation} 
added to Hamiltonian~(\ref{eq:i4}):
\begin{equation}
  {\bf H}(h_{3},\alpha) \equiv {\bf H}(h_{3})+
  \alpha {\bf H}_{2},
  \label{eq:i13}
\end{equation}
does not change the position of the soft modes $q_1=\pi$ and $q_3(M)$ but
changes the associated $\eta_1(M,\alpha),\;\eta_3(M,\alpha)$
exponents.\cite{GFA+97} A singlet triplet gap opens in the free field case
$(h_{3}=0)$ for $\alpha\!>\!\alpha_c\!=\!0.241\ldots$.\cite{ON92} Note, however,
that (\ref{eq:i12}) is translation invariant and therefore the ground state
degeneracy with momenta $p=0,\pi$ -- predicted by the LSM construction -- still
holds, i.e. the singlet ground state is still twofold degenerate in the singlet
sector.
%%%%%%%%%%%%%%%%%%%%%%%%%%%%%%%%%%%%%%%%%%%%%%%%%%%%%%%%%%%%%%%%%%%%%%%%%%%%%%%%  
\subsection{A staggered dimer field}\label{sec:OalphaDpi}
%%%%%%%%%%%%%%%%%%%%%%%%%%%%%%%%%%%%%%%%%%%%%%%%%%%%%%%%%%%%%%%%%%%%%%%%%%%%%%%%  
A plateau in the magnetization curve at $M=1/4$ has been found in the
Hamiltonian (\ref{eq:i4}) with an additional next-to-nearest-neighbor coupling
and a staggered dimer field: \cite{TNK98,Tots98}
\begin{equation}
  {\bf H}(h_{3},\alpha,\delta) \equiv {\bf H}(h_{3})+
  \alpha {\bf H}_{2} + \delta  {\bf D}(\pi).
  \label{eq:d1}
\end{equation}
The dimer operator is defined as:
\begin{equation}
  {\bf D}(q) \equiv 2\sum_{l=1}^{N} e^{iql} {\bf S}_{l} \cdot {\bf S}_{l+1}.
  \label{eq:i14}
\end{equation}
Such a Hamiltonian only commutes with ${\bf T}^2$ and therefore reduces the
degeneracy of the ground state. At $M=0$ the twofold degeneracy of the ground
state is lifted and a gap opens between the energies of the ground state and the
excited states.\cite{FKM98a}
  
At $M=1/4$ a gap opens between the ground state $|0\rangle$ and one further
state, which is even under ${\bf T}^2$. The LSM construction yields a second
state $|1\rangle={\bf U}|0\rangle$ degenerate with the ground state $|0\rangle$,
which is odd under ${\bf T}^2$.
%%%%%%%%%%%%%%%%%%%%%%%%%%%%%%%%%%%%%%%%%%%%%%%%%%%%%%%%%%%%%%%%%%%%%%%%%%%%%%%%  
\subsection{A periodic dimer field}\label{sec:bDq}
%%%%%%%%%%%%%%%%%%%%%%%%%%%%%%%%%%%%%%%%%%%%%%%%%%%%%%%%%%%%%%%%%%%%%%%%%%%%%%%% 
A plateau in the magnetization curve at $M=1/6$ has been found \cite{Hida94}
for a Hamiltonian of the type (\ref{eq:i4}) with a dimer field ${\bf \bar D}(q)$
of period $q=2\pi/3$:
\begin{equation}
  {\bf \bar D}(q) \equiv [{\bf D}(q)+{\bf D}(-q)]/2.
  \label{eq:i14b}
\end{equation}

The Hamiltonian used in Ref.\onlinecite{Hida94} can be reformulated as a single
spin-1/2 chain with ferromagnetic nearest neighbor coupling being strongly
disturbed by an antiferromagnetic next-to-nearest neighbor coupling with a
period of $q=2\pi/3$.

Note, that the periodicity $q=2\pi/3$ coincides with the soft mode momentum
$q_{3}(M=1/6)$. Such a coincidence occurs in both examples~\ref{sec:bS3q}
and~\ref{sec:bDq} and we conclude that the special type of the periodic
perturbation~(\ref{eq:i10}) and~(\ref{eq:i14b}) is not relevant for the
formation of a magnetization plateau. 

The situation in example~\ref{sec:OalphaDpi} is different. Here the periodicity
($q=\pi$) of the staggered dimer field coincides with the {\it second soft mode}
$2q_{3}(M=1/4)=\pi$ [~(\ref{eq:i7}) for $k=2$], predicted by the LSM
construction. Note, however, that the magnetization plateau at $M=1/4$ is only
visible if the parameters $\alpha$ and $\delta$ in~(\ref{eq:d1}) are
appropriately chosen. In particular the magnetization plateau at $M=1/4$ seems
to be absent if the next-to-nearest-neighbor coupling $\alpha$ is switched off.
Therefore, we conclude that the coincidence of the periodicity $q$ in the
perturbation operator with the momentum of one LSM-soft mode is a necessary --
but not sufficient -- condition for the formation of a plateau.

It is the purpose of this paper to investigate in more detail the mechanism for
the formation of gaps and magnetization plateaus by means of periodic
dimer-perturbations~(\ref{eq:i14b}) of strength $2\delta_{q}$. The
$\delta_{q}$-evolution of the energy eigenvalues and transition matrix elements
of the perturbation operator is given by a closed set of differential equations,
which we have discussed in Refs.\onlinecite{FKM98a,FKM98b}. The initial values for
these evolution equations are given by the energy eigenvalues and transition
amplitudes for the unperturbed case ($\delta_{q}=0$).

The outline of the paper is as follows. In Sec.~\ref{sec:LSM-construction} we
complete the discussion of the quantum numbers of the LSM state $|k\rangle$ by
investigating their ${\bf S}_{T}^{2}$ content, where ${\bf S}_{T}$ is the total
spin operator. We are in particular interested in the question, whether or not
the LSM state $|1\rangle$ at $M=0$ with momentum $p_{1}=p_{0}+\pi$ contains a
triplet ($S=1$) or higher spin component [${\bf S}_{T}^{2}=S(S+1)$].
Section~\ref{sec:sm-dimer} is devoted to an analysis of the LSM soft modes in
the dimer-dimer structure factor. This analysis is used to fix the above
mentioned initial conditions for the evolution equations. In
Sec.~\ref{sec:gaps-plateau} we then present numerical results on the formation
of gaps and plateaus by means of the periodic dimer perturbations.

The occurrence of magnetization plateaus in spin ladders is discussed in
Sec.~\ref{sec:spin-ladders}.  
%%%%%%%%%%%%%%%%%%%%%%%%%%%%%%%%%%%%%%%%%%%%%%%%%%%%%%%%%%%%%%%%%%%%%%%%%%%%%%%%
%
\section{The Lieb, Schultz, Mattis (LSM) Construction and the quantum numbers 
         of the degenerate ground states.}\label{sec:LSM-construction}
%
%%%%%%%%%%%%%%%%%%%%%%%%%%%%%%%%%%%%%%%%%%%%%%%%%%%%%%%%%%%%%%%%%%%%%%%%%%%%%%%%
It has been pointed out in the introduction that the quantum number
analysis of the states $|k\rangle={\bf U}^k|0\rangle$ in the LSM
construction is crucial to decide whether these states are new, i.e.
orthogonal to the ground state $|0\rangle$, or not. The transformation
behavior (\ref{eq:i6}) under translations ${\bf T}$ yields the momenta
$p_k$ (\ref{eq:i7}) of these states.
       
The operator ${\bf U}$ (\ref{eq:i1}) obviously commutes with the total spin in
3-direction $S^3_T$.  Therefore all the states $|k\rangle = {\bf U}^k |0\rangle$
have the same total Spin $S^3_T$ in 3-direction.  The Hamiltonian of
type~(\ref{eq:i13}) with isotropic couplings commutes with ${\bf S}^2_T$. One
might ask for the ${\bf S}_T^{2}$ content of the states $|k\rangle$. To answer
this question we compute the expectation value
\begin{eqnarray}
  \langle k| {\bf S}^2_T |k \rangle - \langle 0|{\bf S}^2_T |0 \rangle &=&
  \langle 0| {\bf U}^{\dagger k}{\bf S}^2_T {\bf U}^k |0\rangle -
  \langle0|{\bf S}^2_T |0 \rangle \nonumber\\ && \hspace*{-2cm}
  = 2 N [{S}_1(q=k2\pi/N,M)-{S}_1(0,M)],
\label{eq:lsm1}
\end{eqnarray}
using the considerations developed by LSM to show the vanishing of the energy
difference (\ref{eq:i3}). The right-hand side of (\ref{eq:lsm1}) is determined
by the transverse structure factor, exposed to an external field $h_{3}$ with
magnetization $M(h_{3})$:
\begin{equation}
  {S}_1(q,M) \equiv \sum_{l=1}^{N} e^{iql} 
  \langle S,p_{s}|S^{1}_{1}S^{1}_{1+l}|S,p_{s} \rangle ,
  \label{eq:lsm2}
\end{equation}
which has been studied on finite systems in Ref.~\onlinecite{KMS94} for the
nearest neighbor model (\ref{eq:i4}) and in Ref.~\onlinecite{SGMK96} for the
model with next-to-nearest-neighbor couplings $\alpha$. 

From these investigations we conclude for the thermodynamical limit of the
difference appearing on the right-hand side of (\ref{eq:lsm1}):
\begin{equation}
  2N[{S}_1(k2\pi/N,M)\!-\! {S}_1(0,M)] 
  \stackrel{N\to\infty}{\longrightarrow}
  A(M)\left(\frac{2\pi}{N}\right)^{\beta_{k}}.
\label{eq:lsm3}
\end{equation}
The exponent $\beta_{k}=\beta_{k}(M,\alpha)$ turns out to be zero for $M=0$ and
$\alpha=0$ [cf.  Fig.~5(d) in Ref.~\onlinecite{KMS94}].  This means, that the
right-hand side of (\ref{eq:lsm1}) is non vanishing. For $M=0,\alpha=0$ the
ground state is a singlet state [${\bf S}^2_T = S(S+1)=0$] and (\ref{eq:lsm1})
tells us that the soft mode state $|k=1\rangle = {\bf U}|0\rangle$ with momentum
$\pi$ contains triplet $[{\bf S}^2_T=S(S+1)=2]$ and higher spin components, i.e.
the LSM construction together with (\ref{eq:lsm3}) and
$\beta_{k}(M=0,\alpha=0)=0$ forbids a singlet triplet gap.

The exponent $\beta_{k}(M,\alpha)$ is larger than zero for $M>0$ [cf. Inset of
Fig.~5(b) in Ref.~\onlinecite{KMS95}] In this case the right-hand side of
(\ref{eq:lsm1}) vanishes and the soft mode states $|k\rangle = {\bf U}^k
|0\rangle$ have the same total spin ${\bf S}^2_T = S(S+1)$ as the ground state.

Switching on the next-to-nearest-neighbor coupling $\alpha$ the free field
exponent must be larger than zero
\begin{equation}
  \beta_{1}(0,\alpha) >  0 \quad \mbox{for} \quad \alpha > \alpha_c=0.241\ldots,
\label{eq:lsm6}
\end{equation}
since the dimer phase $\alpha > \alpha_c$ is characterized by a singlet triplet
gap.\cite{ON92} In other words, for $\alpha > \alpha_c$ the degenerate LSM state
$|1\rangle = {\bf U}|0\rangle$ with momentum $p_{s}+\pi$ must be a pure singlet
state as well. The exponent $\beta_{1}(0,1/2)$ can easily be calculated at the
Majumdar-Ghosh \cite{MG69,MG69b} point $\alpha=1/2$ . Here we find
$\beta_{1}(0,1/2)=1$.

\begin{figure}[ht]
\centerline{\epsfig{file=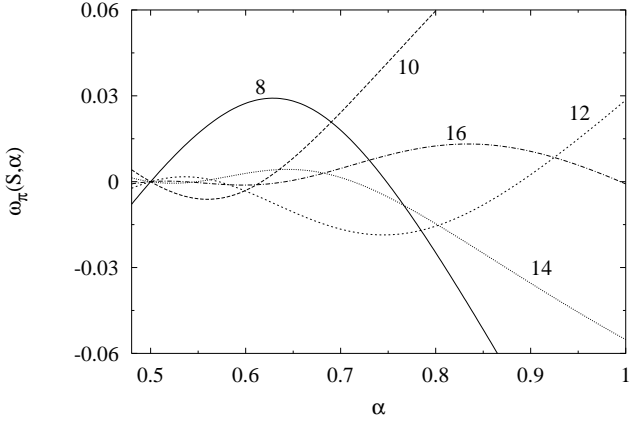,width=6.0cm,angle=-90}}
\caption{The energy differences~(\ref{eq:lsm7}) for system sizes
  $N=8,10,\ldots,16$ and $S=0$.}
\label{fig:plot2a}
\end{figure}

In Fig.~\ref{fig:plot2a} we have plotted the energy difference $\omega_{\pi}
= E_1 - E_0$ of the two singlet states $|0\rangle$ and
$|1\rangle$
%%%%%%%%%%%%%%%%%%%%%%%%%%%%%%%BEGIN-FIGURE%%%%%%%%%%%%%%%%%%%%%%%%%%%%%%%%
%%%%%%%%%%%%%%%%%%%%%%%%%%%%%%%%END-FIGURE%%%%%%%%%%%%%%%%%%%%%%%%%%%%%%%%%
\begin{equation}
\omega_{\pi}(S=0,\alpha) = E(p_{s}+\pi,S=0,\alpha) - E(p_{s},S=0,\alpha),
\label{eq:lsm7}
\end{equation}
versus the coupling $\alpha$. This is an oscillating function for $\alpha > 0.5$
with zeroes at:
\begin{equation}
\alpha = \alpha_1(N) < \alpha_2(N) < ... < \alpha_Z(N),
\label{eq:lsm8}
\end{equation}
where numerical data suggest that the total number of zeroes is given by
\begin{equation}
  Z = \frac{1}{4} 
  \begin{cases}
    N &: N=8,12,16,\ldots \\
  N-2 &: N=10,14,18,\ldots  .
  \end{cases}
\label{eq:lsm9}
\end{equation}
In the thermodynamical limit we have a dense distribution of zeroes. The height
of the maxima and minima in between converges to zero -- a signal for the
degeneracy of the two states $|0\rangle$ and $|1\rangle = {\bf U} |0\rangle$ in
the thermodynamical limit as predicted by the LSM construction.

%%%%%%%%%%%%%%%%%%%%%%%%%%%%%%%BEGIN-FIGURE%%%%%%%%%%%%%%%%%%%%%%%%%%%%%%%%
\begin{figure}[ht]
\centerline{\epsfig{file=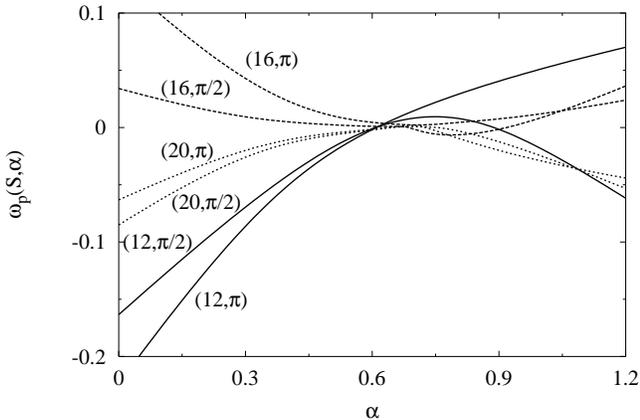,width=6.0cm,angle=-90}}
\caption{The energy differences~(\ref{eq:lsm10}) for system sizes
  $N=8,12,16$ and momentum $p=\pi,\pi/2$ and $S=N/4$. Same line types denote the
  same system sizes.}
\label{fig:plot2b}
\end{figure}
%%%%%%%%%%%%%%%%%%%%%%%%%%%%%%%%END-FIGURE%%%%%%%%%%%%%%%%%%%%%%%%%%%%%%%%%
As was pointed out in the introduction, a fourfold degeneracy of the states
$|k\rangle = {\bf U}^k|0\rangle,\; (k=0,1,2,3) $ is predicted at $M=1/4$. All
these states have the same total spin squared ($S=N/4$) and momenta $p=0,\pm
\pi/2,\pi$. On finite systems one again observes oscillations in the energy
differences
\begin{equation}
\omega_p(S,\alpha) \equiv E(p,S=N/4,\alpha) - E(0,S=N/4,\alpha),
\label{eq:lsm10}
\end{equation}
for $p=\pi/2,\pi$, if we switch on the next-to-nearest-neighbor coupling
$\alpha$ as is shown in Fig.~\ref{fig:plot2b}.  For $\alpha$ values large enough
these oscillations die out and the ground state momentum $p_{s}, S=N/4 $ is
supposed to be
\begin{equation}
p_{s}(1/4)=\frac{\pi}{2}
\begin{cases}
   2 &: N=8,24,40,\ldots \\ 
   1 &: N=12,28,44,\ldots \\ 
   0 &: N=16,32,48,\ldots \\ 
   1 &: N=20,36,52,\ldots.
\end{cases}
\label{eq:lsm11}
\end{equation}
%%%%%%%%%%%%%%%%%%%%%%%%%%%%%%%%%%%%%%%%%%%%%%%%%%%%%%%%%%%%%%%%%%%%%%%%%%%%%%%%
%
\section{soft modes in the dimer dimer structure factor}\label{sec:sm-dimer}
\begin{figure}[ht]
\centerline{\epsfig{file=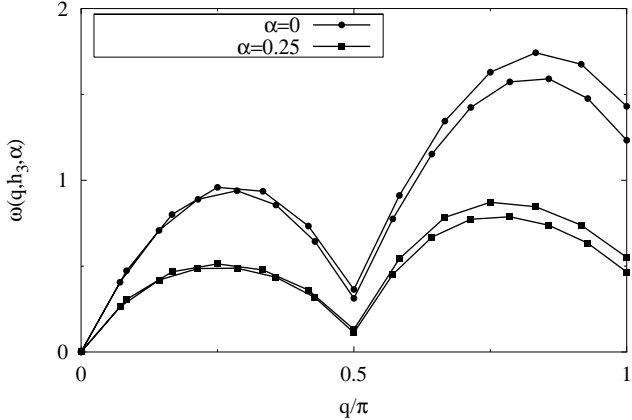,width=6.0cm,angle=-90}}
\caption{The dispersion curve of the Hamiltonian~(\ref{eq:i13}) for finite
  system ($N=24,28$) and magnetization $M=1/4$.}
\label{fig:disp}
\end{figure}

%
%%%%%%%%%%%%%%%%%%%%%%%%%%%%%%%%%%%%%%%%%%%%%%%%%%%%%%%%%%%%%%%%%%%%%%%%%%%%%%%%
According to the LSM construction for the translation invariant models with
nearest and next-to-nearest-neighbor couplings (\ref{eq:i13}), ${\bf
H}(h_{3},\alpha)$, we expect the dispersion curve
\begin{equation}
  \omega_{q}(S,\alpha) \equiv E(p_{s}+q,S,\alpha) - E(p_{s},S,\alpha),
\label{eq:31}
\end{equation}
to develop zeroes
\begin{equation}
\omega^{(k)}(h_{3},\alpha)\equiv \omega_{q}(S,\alpha), 
\label{eq:32}
\end{equation}
at the soft mode momenta $q=q^{(k)}(M) \equiv k \pi (1-2M)$. If in the
thermodynamical limit the scaled energy differences: 
\begin{equation}
  \hat\Omega^{(k)}(M,\alpha) \equiv \lim_{N\to\infty} N\omega^{(k)}(h_{3},\alpha).
\label{eq:32b}
\end{equation}
and 
\begin{equation}
  v(M,\alpha) = \lim_{N \to \infty} N 
      [E(p_{s}\!+\!2\pi/N,S,\alpha) - E(p_{s},S,\alpha)],
\label{eq:34}
\end{equation}
are finite and non vanishing, the ratios
\begin{equation}
\eta^{(k)} (M,\alpha) \equiv \
     \frac{\hat\Omega^{(k)} (M,\alpha)}{\pi v(M,\alpha)},
\label{eq:33}
\end{equation}
yield the exponents $\eta^{(k)}(M,\alpha)$, which govern the critical behavior
of the dimer dimer structure factor: 
\begin{equation}
  S_{DD} (q,M) = \frac{1}{N} 
    \langle p_{s},S|{\bf D}_c(q) {\bf D}_c^\dag (q) |p_{s},S \rangle.
\label{eq:35}
\end{equation}
Here 
\begin{displaymath}
{\bf D}_c(q) \equiv {\bf D}(q) - \delta_{q0} {\bf D}(0)
\end{displaymath}
is the connected part of the dimer operator~(\ref{eq:i14}).  
For $M=0$ a zero should occur at $q=\pi$ and for $M=1/4$ we should find two
zeroes at $q=\pi/2$ and $q=\pi$.  This is indeed the case as can be seen from
Fig.~\ref{fig:disp} where we have plotted the dispersion curves for
$M=1/4$, $\alpha=0,1/4$.
%%%%%%%%%%%%%%%%%%%%%%%%%%%%%%%BEGIN-FIGURE%%%%%%%%%%%%%%%%%%%%%%%%%%%%%%%%
%%%%%%%%%%%%%%%%%%%%%%%%%%%%%%%%END-FIGURE%%%%%%%%%%%%%%%%%%%%%%%%%%%%%%%%%
Note, that the dimer operator commutes with the total Spin ${\bf S}^2_T$ and the
dispersion curve~(\ref{eq:31}) describes the lowest lying excitations
contributing to $S_{DD}(q,M,\alpha)$.

\begin{figure}[ht]
\centerline{\epsfig{file=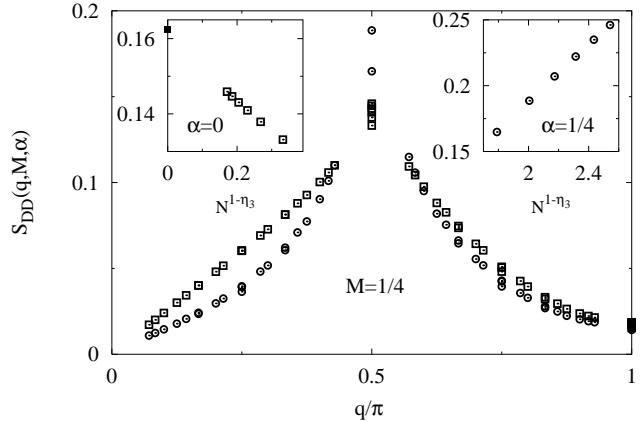,width=6cm,angle=-90}}
\caption{The dimer dimer structure factor~(\ref{eq:35b}) at $M=1/4$ for
  $N=12,16,\ldots,28$. The insets show the $N$-dependence of the structure
  factor $S_{DD}$ at $q=\pi/2$, versus $N^{1-\eta_{3}}$, with exponent
  $\eta_{3}$ given by Eq.~(\ref{eq:eta_a_M1d4}).  For $\alpha=1/4$ the structure
  factor diverges and for $\alpha=0$ it is finite.  The extrapolated value
  ($N\to\infty$) is marked with a solid symbol ($\blacksquare$). }
\label{fig:SDD_q}
\end{figure}

The $q$-dependence of the dimer-dimer structure factor~(\ref{eq:35}) at $M=1/4$
and $\alpha=0,1/4$ is shown in Fig.~\ref{fig:SDD_q}. A pronounced peak is found
at the first soft mode $q=\pi(1-2M)=\pi/2$. The finite-size behavior of the
structure factor at the first soft mode 
\begin{equation}
  S_{DD} (q_{3}(M),M,\alpha) \sim N^{1-\eta_{3}(M,\alpha)}
\label{eq:37}
\end{equation}
is shown for $\alpha=0,1/4$ and $M=1/4$ in the inset of Fig.~\ref{fig:SDD_q}.
It is well described by an exponent
\begin{equation}\label{eq:eta_a_M1d4}
\eta^{(1)}(1/4,\alpha) = \eta_{3}(1/4,\alpha) = 
  \begin{cases}
     1.53 &: \alpha=0\\ 
     0.72 &: \alpha=1/4.
  \end{cases}
\end{equation}
identical with the exponent $\eta_{3}(M,\alpha)$ in the longitudinal structure
factor. The latter has been calculated exactly in the model with nearest
neighbor couplings ($\alpha=0$) by means of Bethe ansatz solutions for the
energy eigenvalues, which enter in the differences~(\ref{eq:31})
and~(\ref{eq:34}). The resulting curve $\eta_{3}(M,\alpha= 0)$ is shown in
Fig.~2 of Ref.\onlinecite{FGM+96}. The $\alpha$-dependence of
$\eta_{3}(1/4,\alpha)$ has been calculated on small systems $(N\leq 28)$ and can
be seen in Fig.~4(b) of Ref.\onlinecite{GFA+97}.

According to its definition~(\ref{eq:35}), the dimer dimer structure factor can
be represented:
\begin{equation}
  S_{DD} (q,M,\alpha) \equiv \frac{1}{N} 
  \sum_{n} \left| \langle n | {\bf D}_{c}(q) | 0 \rangle \right|^{2}
\label{eq:35b}
\end{equation}
in terms of transition amplitudes from the ground state $|0\rangle =|
p_{s},S\rangle$ to excited states $|n\rangle$ with momenta $p_{n}=p_{s}+q$ and
total spin $S$.  The peak in Fig.\ref{fig:SDD_q} tells us that the transition
matrix elements at the first soft mode $q=\pi(1-2M)$
\begin{equation}
  \label{eq:13}
  \langle n|{\bf D}(\pi(1-2M))|0\rangle 
  \stackrel{N\to\infty}{\longrightarrow}
  N^{\kappa_{3}},
\end{equation}
with $\kappa_{3}=1-\eta_{3}(M(h_{3}),\alpha)/2$, diverge with the
system size. 
%%%%%%%%%%%%%%%%%%%%%%%%%%%%%%%BEGIN-FIGURE%%%%%%%%%%%%%%%%%%%%%%%%%%%%%%%%
%%%%%%%%%%%%%%%%%%%%%%%%%%%%%%%%END-FIGURE%%%%%%%%%%%%%%%%%%%%%%%%%%%%%%%%%
The fact, that there is no peak at the second soft mode, indicates that the
corresponding transition matrix elements $\langle n | \bar {\bf D}(q=2\pi(1-2M))
| 0 \rangle$ are small -- at least for next-to-nearest-neighbor couplings
$\alpha\leq 1/4$. The magnitude of these transition matrix elements is crucial
for the formation of gaps and magnetization plateaus with a periodic
perturbation $\bar {\bf D}(q)$.
%%%%%%%%%%%%%%%%%%%%%%%%%%%%%%%%%%%%%%%%%%%%%%%%%%%%%%%%%%%%%%%%%%%%%%%%%%%%%%%%
%
\section{Periodic dimer perturbations and the formation of gaps and plateaus}
\label{sec:gaps-plateau}
%
%%%%%%%%%%%%%%%%%%%%%%%%%%%%%%%%%%%%%%%%%%%%%%%%%%%%%%%%%%%%%%%%%%%%%%%%%%%%%%%%
In this section we will study the impact of a periodic dimer perturbation
${\bf \bar D}(q)$~(\ref{eq:i14b}):
\begin{equation}
  {\bf H}(h_{3},\alpha,\delta_{q}) \equiv 
   {\bf H}(h_{3},\alpha)+ 2\delta_{q}  {\bf \bar D}(q),
  \label{eq:Hh3deltaq}
\end{equation}
on the soft modes. We will restrict ourselves to the model with nearest neighbor
and next-to-nearest-neighbor couplings and follow the procedure of
Refs.~\onlinecite{FKM98a,FKM98b}.  There it was shown that the
$\delta_{q}$-evolution of the energy eigenvalues and transition matrix elements
$\langle n | \bar{\bf D}(q)|0\rangle$ is described by a closed set of
differential equations which possess finite size scaling solutions in the limit
$N \to \infty$, $\delta_{q} \to 0$, $x=\delta_{q}^{\epsilon} N $ fixed.

\begin{figure}[htb]
\centerline{\epsfig{file=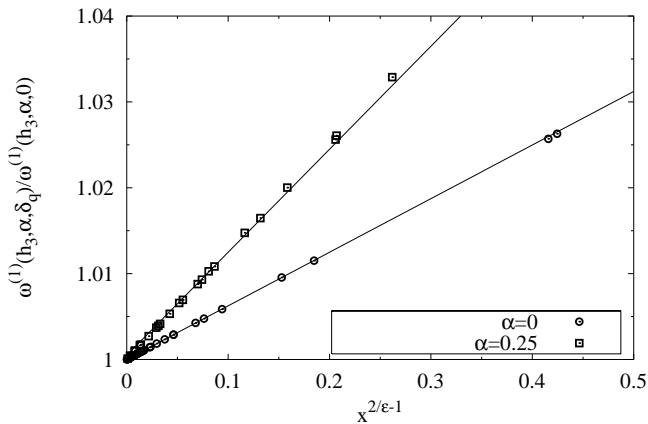,width=6.0cm,angle=-90}}
\caption{The gap ratio~(\ref{eq:45}) versus the scaling variable
  $(N\delta^{\epsilon})^{2/\epsilon-1}$, with $\epsilon$ given by
  Eq.~(\ref{eq:46}).  The solid lines represent linear fits to the small
  $x$-behavior.}
\label{fig:plot4a}
\end{figure}

This statement holds, if the periodicity $q$ of the dimer perturbation coincides
with a soft mode momentum:
\begin{equation}
  q = q^{(k)}(M) = k \pi (1-2M).
\label{eq:42}
\end{equation}
Then a gap in the energy difference (\ref{eq:31}) between the ground state and
lowest state which can be reached with the perturbation $\bar{\bf D}(q)$ at
$q^{(k)}(M)$ is predicted:
\begin{equation}
  \omega^{(k)} (h_{3},\alpha,\delta_{q}) = 
  \delta_{q}^{\epsilon^{(k)}} \Omega^{(k)} (M,\alpha,x).
  \label{eq:43}
\end{equation}
The gap  opens with an exponent $\epsilon^{(k)}=\epsilon^{(k)}(h_{3},\alpha)$,
related to the corresponding $\eta$-exponent (\ref{eq:33}) via
\begin{equation}
  \epsilon^{(k)}(h_{3},\alpha) = 2[4-\eta^{(k)}(M(h_{3}),\alpha)]^{-1}.
\label{eq:44}
\end{equation}
A test of the prediction (\ref{eq:43}) is given in Fig.~\ref{fig:plot4a}, where
we plotted the gap ratio
\begin{equation}
  \frac{\omega^{(1)}(h_{3},\alpha,\delta_{q})}{\omega^{(1)}(h_{3},\alpha,0)}
       = 1 + e^{(1)}(h_{3},\alpha,x),
\label{eq:45}
\end{equation}
for $M=1/4$, i.e. $q=\pi/2$ and $\alpha=0,1/4$ versus the scaling variable
$x^{2/\epsilon}$ with the exponents
\begin{eqnarray}\epsilon = \epsilon^{(1)} (1/4,\alpha) =
  \begin{cases}
   0.81(1)&: \alpha=0 \\
   0.64(3)&: \alpha=1/4 
  \end{cases}
\label{eq:46}
\end{eqnarray}
%%%%%%%%%%%%%%%%%%%%%%%%%%%%%%%BEGIN-FIGURE%%%%%%%%%%%%%%%%%%%%%%%%%%%%%%%%
%%%%%%%%%%%%%%%%%%%%%%%%%%%%%%%%END-FIGURE%%%%%%%%%%%%%%%%%%%%%%%%%%%%%%%%%

Note that the gap ratio (\ref{eq:45}) is linear in $x^{2/\epsilon}$ for small
values of $x$, as predicted by the evolution equations for the scaling
functions.\cite{FKM98a,FKM98b} 

Let us next study the influence of the periodic perturbation~(\ref{eq:i14b}) on
the magnetization curve $M=M(h_{3})$. A plateau in the magnetization curve with
an upper and lower critical field $h_{3}^{u}$ and $h^{l}_{3}$ will emerge if
\begin{eqnarray}
\Delta({\delta_{q}},h_{3})
    \equiv h_{3}^{u}-h_{3}^{l} &=& \lim_{N \to \infty} \left[ 
  E(p_{s\!+\!1},S+1,\alpha,\delta_{q})\right.  \nonumber \\  && \hspace{-1.5cm}
 \left. \hspace{-1cm} + E(p_{s\!-\!1},S-1,\alpha,\delta_{q}) - 2
  E(p_{s},S,\alpha,\delta_{q})
 \right],
\label{eq:47}
\end{eqnarray}
does not vanish in the thermodynamical limit, i.e. if the ground state energies
$E(p_{s'},S',\alpha,\delta_{q})$ for $S'=S-1,S,S+1$ evolve in a different manner
under the perturbation $\delta_{q}$. This happens exactly if the periodicity $q$ of
the dimer perturbation coincides with a soft mode momentum (\ref{eq:42}). Then
the ground state energy at the 'critical' magnetization $M=S_{T}^{3}/N$ is
lowered stronger than at the neighboring magnetizations $M=(S_{T}^{3}\pm1)/N$.

\begin{figure}[ht]
\vspace{5cm}
\centerline{\hspace{4cm}\epsfig{file=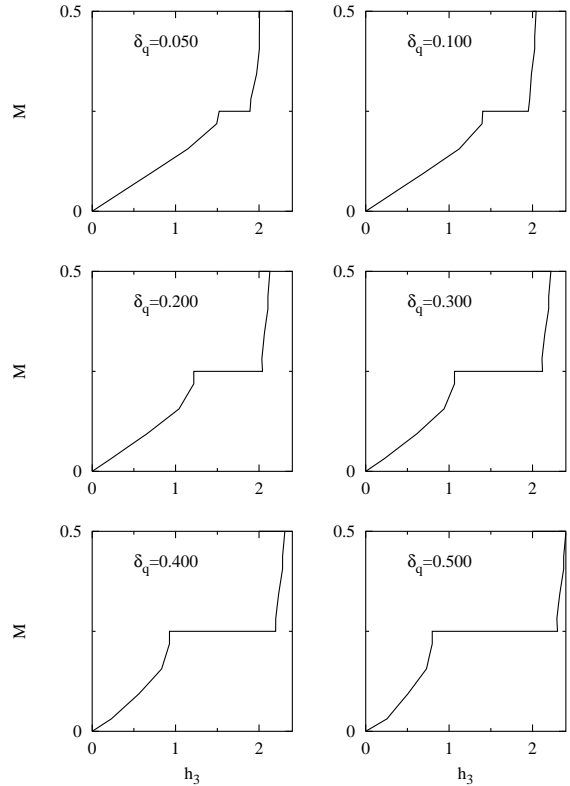,width=4cm}}
\caption{The magnetization curve of ${\bf
    H}(h_{3},\alpha,\delta_{q})$ for $\alpha=0.25$ and different
  $\delta_{q}$-values, determined from finite system sizes $N=8,12,16,20$. }
\label{fig:fig_plateau_a0.25}
\end{figure}

In Fig.~\ref{fig:fig_plateau_a0.25} we show the $\delta_{q}$-evolution of the
magnetization curves for $q=\pi/2$ and $\alpha=1/4$. The emergence of the
predicted plateau at $M=1/4$ is clearly visible.

%%%%%%%%%%%%%%%%%%%%%%%%%%%%%%%BEGIN-FIGURE%%%%%%%%%%%%%%%%%%%%%%%%%%%%%%%%
%%%%%%%%%%%%%%%%%%%%%%%%%%%%%%%%END-FIGURE%%%%%%%%%%%%%%%%%%%%%%%%%%%%%%%%%

\begin{figure}[hvt]
\centerline{\epsfig{file=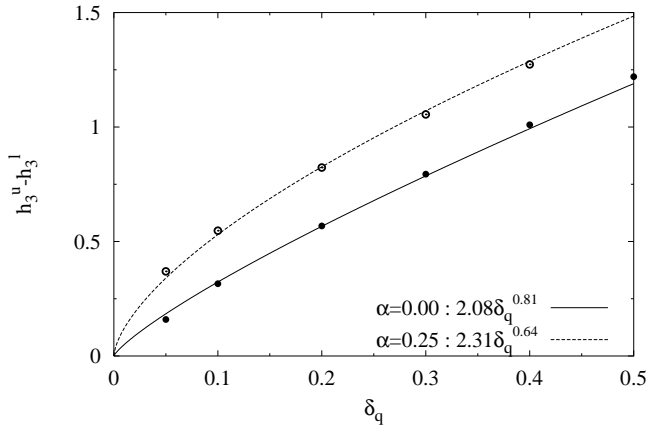,width=6cm,angle=-90}}
\caption{The evolution of the difference~(\ref{eq:47}) between the upper and
  lower critical field at the plateau $M=1/4$. The dotted and dashed line show a
  fit, proportional to $\delta_{q}^{\epsilon}$ with $\epsilon=\epsilon^{(1)}$
  given by Eq.~(\ref{eq:46}), to small values of the external field
  $\delta_{q}$.}
\label{fig:fig_hu-hl_m1d4}
\end{figure}

The scaling behavior $\delta_{q}^{\epsilon}$ with
$\epsilon=\epsilon^{(1)}(h_{3},\alpha)$ of the difference (\ref{eq:47}) is
governed by the critical exponent $\eta^{(1)}(h_{3},\alpha)$ at
$h_{3}=h_{3}(M=1/4)$ of the unperturbed model at the first soft mode $q_{3}(M)$
as can be seen in Fig.~\ref{fig:fig_hu-hl_m1d4}.

We also looked for the $\delta$-evolution of the magnetization curves for a
dimer perturbation with periodicity $q=\pi$. There is a plateau for $M=0$ --
corresponding to the gap above the ground state discussed in
Ref.~\onlinecite{GFA+97}. However, no plateau is visible
at $M=1/4$ for small perturbations $\delta {\bf D}(\pi)$. This corresponds to
the observation that the second soft mode at $M=1/4$, $q=\pi$, does not produce
any signature in the dimer dimer structure factor.

These statements only hold for small perturbations $\delta {\bf D}(\pi)$. It is
indeed known from Refs.~\onlinecite{TNK98,Tots98} that a plateau at $M=1/4$ can
be enforced with a large perturbation of order $1$.

%%%%%%%%%%%%%%%%%%%%%%%%%%%%%%%BEGIN-FIGURE%%%%%%%%%%%%%%%%%%%%%%%%%%%%%%%%
%%%%%%%%%%%%%%%%%%%%%%%%%%%%%%%%END-FIGURE%%%%%%%%%%%%%%%%%%%%%%%%%%%%%%%%%

%%%%%%%%%%%%%%%%%%%%%%%%%%%%%%%BEGIN-FIGURE%%%%%%%%%%%%%%%%%%%%%%%%%%%%%%%%
\begin{figure}[htb]
\vspace{5cm}
\centerline{\hspace{4cm}\epsfig{file=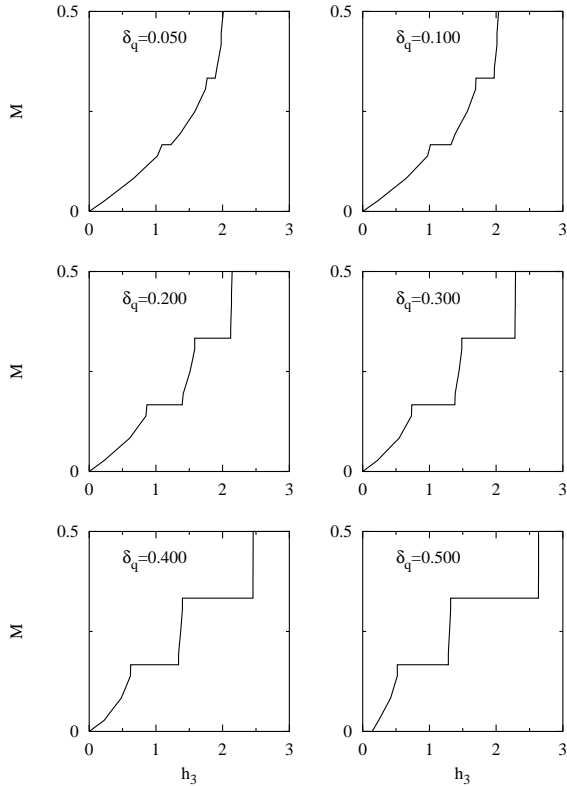,width=4cm}}
\caption{The magnetization curve of Hamiltonian~(\ref{eq:Hh3deltaq}) with a
  perturbation $\bar {\bf D}(2\pi/3) + \bar {\bf D}(\pi/3)$ for $\alpha=0$
  and different $\delta_{q}$-values, determined from finite system sizes
  $N=6,12,18$.}
\label{fig:fig_plateau_2}
\end{figure}
%%%%%%%%%%%%%%%%%%%%%%%%%%%%%%%%END-FIGURE%%%%%%%%%%%%%%%%%%%%%%%%%%%%%%%%%

We have computed magnetization curves for the Hamiltonian~(\ref{eq:Hh3deltaq})
with perturbations $\bar {\bf D}(q)$ of period $q=2\pi/3$ and $q=\pi/3$. We
found clear evidence for the expected magnetization plateaus at $M=1/6$ and
$M=1/3$. The magnetization curve for a Hamiltonian with a superposition of both
perturbations
\begin{equation}
  \label{eq:11}
  \bar{\bf D}(2\pi/3) + \bar{\bf D}(\pi/3),
\end{equation}
is shown in Fig.~\ref{fig:fig_plateau_2}. Here, we find two plateaus in the
magnetization curve at $M=1/6$ and $M=1/3$.
%%%%%%%%%%%%%%%%%%%%%%%%%%%%%%%%%%%%%%%%%%%%%%%%%%%%%%%%%%%%%%%%%%%%%%%%%%%%%%%%
%
\section{Magnetization plateaus in spin ladders}\label{sec:spin-ladders}
\begin{figure}[ht]
\centerline{\epsfig{file=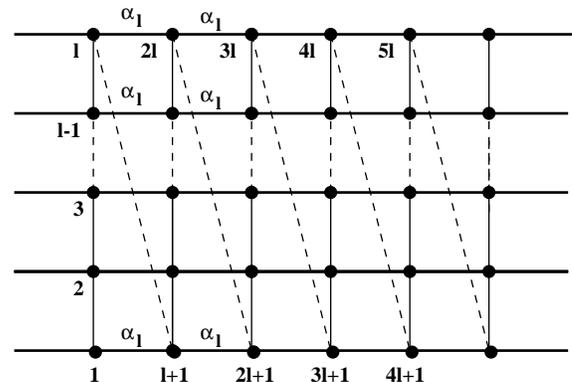,width=7.5cm}}
\caption{A spin ladder system described by Hamiltonian~(\ref{eq:H_h3_al}) 
  with $l$ legs, and additional diagonal couplings (dashed lines). The coupling
  strength between the legs is one unit.}
\label{fig:l_ladder}
\end{figure}

%
%%%%%%%%%%%%%%%%%%%%%%%%%%%%%%%%%%%%%%%%%%%%%%%%%%%%%%%%%%%%%%%%%%%%%%%%%%%%%%%%
All the considerations we made so far for the 1D
Hamiltonian~(\ref{eq:Hh3deltaq}) with nearest and next-to-nearest-neighbor
couplings can be extended to the case
\begin{equation}
  \label{eq:H_h3_al}
  {\bf H}(h_{3},\alpha_{l}) \equiv {\bf H}(h_{3}) + \alpha_{l} {\bf H}_{l},
\end{equation}
where we substitute the next-to-nearest-neighbor coupling by couplings over $l$
lattice spacings
\begin{equation}
  \label{eq:Hl}
  {\bf H}_{l} \equiv 2 \sum_{n=1}^{N} {\bf S}_{n} \cdot {\bf S}_{n+l}. 
\end{equation}
For $l$ finite the position $q_{3}(M)$ of the first soft mode --
generated by the LSM construction -- will not change, the corresponding
$\eta$-exponent might change. According to our experience with the case
$l=2$, we expect that a slightly frustrating coupling enhances the singularity
in the dimer dimer structure factor at $q=q_{3}(M)$. Hamiltonians of the
type~(\ref{eq:H_h3_al}) are interesting, since they can be viewed as a spin
ladder system with $l$ legs, as is shown in Fig.~\ref{fig:l_ladder}.
%%%%%%%%%%%%%%%%%%%%%%%%%%%%%%%BEGIN-FIGURE%%%%%%%%%%%%%%%%%%%%%%%%%%%%%%%%
%%%%%%%%%%%%%%%%%%%%%%%%%%%%%%%%END-FIGURE%%%%%%%%%%%%%%%%%%%%%%%%%%%%%%%%%
\clearpage
They differ, however, from usual spin ladder systems with $l$ legs, owing to the
diagonal (dashed) couplings [($l\!\leftrightarrow\! l+1),(2l \!\leftrightarrow\!
2l+1),\ldots$], which are inferred by the helical boundary conditions. 

Indeed, these additional couplings change the physical properties. Spin ladder
systems with helical boundary conditions are gapless -- irrespective of the
number of legs. This statement holds if the couplings $\alpha_{l}$ along the
legs is chosen properly, e.g. the two leg system with $l=2$ is gapless for
$\alpha\le\alpha_{c}=0.241\ldots.$ Conventional ladder systems are known to be
gapless for an odd number of legs, but to be gapped for an even number of legs.

This fundamental difference becomes clear, when we add  a
special dimer field to~(\ref{eq:H_h3_al})
\begin{equation}
  \label{eq:dimer_l}
  {\bf D}^{(l)} \equiv \sum_{n=1}^{N}J_{n}^{(l)}  {\bf S}_{n} \cdot {\bf S}_{n+1},
\end{equation}
which only affects the diagonal couplings [($l\leftrightarrow
l+1),(2l\leftrightarrow 2l+1),\ldots$], in Fig.~\ref{fig:l_ladder}:
\begin{eqnarray}
  \label{eq:2}
  J_{n}^{(l)} &=&
  \begin{cases}
    0 &: n=1,\ldots,l-1 \\ \delta &: n=l,
  \end{cases} \\
  \label{eq:14}
    J_{n+l}^{(l)} &=&   J_{n}^{(l)}. 
\end{eqnarray}
The periodicity~(\ref{eq:14}) of the couplings $J_{n}^{(l)}$ is given by a
Fourier series
\begin{equation}
  \label{eq:1}
  J_{n}^{(l)} = \sum_{j=0}^{[l/2]} \cos(2\pi nj/l) \delta(q=2\pi j/l),
\end{equation}
where $[l/2]$ is the largest integer smaller than $l/2$. The Fourier coefficients
$\delta(q)$ follow from~(\ref{eq:2}), e.g. for $l=2$ we find:
\begin{equation}
  \label{eq:3}
  \delta(q=0) = \delta(q=\pi) = \delta/2.
\end{equation}
The dimer perturbation:
\begin{equation}
  \label{eq:4}
  \delta{\bf D}^{(2)} = \frac{\delta}{2}[{\bf D}(0)+{\bf D}(\pi)]
\end{equation}
generates a plateau at $M=(1-q/\pi)/2=0$. Therefore, the gap -- typical for the
two leg ladder -- appears immediately if we switch on the dimer
field~(\ref{eq:dimer_l}), which breaks the translation invariance of the 1D
system. 

Let us next consider the three leg ladder ($l=3$). The Fourier coefficients turn
out to be
\begin{equation}
  \label{eq:5}
  \delta(q=0)=\frac{c_{3}}{1+c_{3}}\delta,\quad
  \delta(q=2\pi/3)=\frac{1}{1+c_{3}}\delta,
\end{equation}
with $c_{3}=\cos(\pi/3)$. The $q=2\pi/3$ component in the dimer
field~(\ref{eq:dimer_l}):
\begin{equation}
  \label{eq:6}
  \delta{\bf D}^{(3)} = \delta(q=0){\bf D}(0)
                      +\delta(q=2\pi/3){\bf D}(2\pi/3),
\end{equation}
generates a plateau at $M=1/6$. This is exactly the plateau found in
Ref.~\onlinecite{CHP97}.

Note, that the Fourier decomposition of $\delta {\bf D}^{(l)}$ contains in
general a $q=\pi$ component for even $l$ and no $q=\pi$ component for odd
$l$. We therefore find a gap at $M=0$ for ladders with an even number of legs but
no gap for ladders with an odd number of legs. 

In summary we can say, the Fourier decomposition of the dimer
field~(\ref{eq:dimer_l}) 
\begin{equation}
  \label{eq:7}
  \delta {\bf D}^{(l)} = \sum_{j=0}^{[l/2]} \delta(q=2\pi j/l) 
                       {\bf D}(q=2\pi j/l),
\end{equation}
tells us, where to expect plateaus in the magnetization curve of a spin ladder
system with $l$ legs. The position of plateaus can be seen in
Table~\ref{tab:legs}.

%%%%%%%%%%%%%%%%%%%%%%%%%%%%%%%%%%%%TABLE%%%%%%%%%%%%%%%%%%%%%%%%%%%%%%%%%%%%%
%\begin{table}[ht]
%  \begin{tabular}{c|cccccc}
%    $l$&2&3&4&5&6 \\ \hline
%    $M$&0&1/6&0;1/4&1/10;3/10&0;1/6;1/3
%  \end{tabular}
%  \caption{The possible position of plateaus for spin ladders~(\ref{eq:H_h3_al})
%    with $l$ legs.}
%  \label{tab:legs}
%\end{table}
%%%%%%%%%%%%%%%%%%%%%%%%%%%%%%%%%%%%TABLE%%%%%%%%%%%%%%%%%%%%%%%%%%%%%%%%%%%%%

The number of plateaus increases with the number of legs and one is tempted to
suggest that in the two dimensional limit $l\to\infty$, the magnetization curve
is again a continuous function. It should be noted, however, that the
LSM-construction with the operator~(\ref{eq:i1}) breaks down in the combined
limit $N\to\infty,\;k=\sqrt{N}$ in the sense that~(\ref{eq:i3}) does not hold.

The magnetic properties of the two dimensional Heisenberg model with helical
boundary conditions at $T=0$ have been studied in Ref.~\onlinecite{YM97}.
Concerning the isotropic model with nearest neighbour coupling, there is no
indication for a plateau in the magnetization curve.

\begin{figure}[ht]
\centerline{\epsfig{file=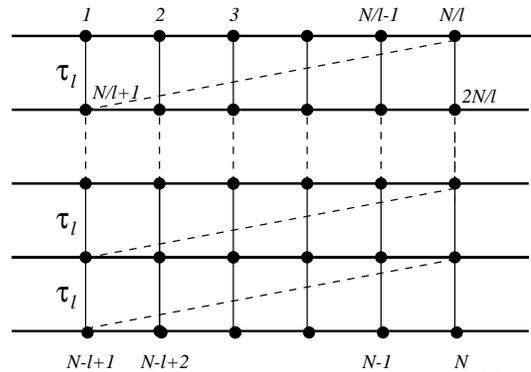,width=7cm}}
\caption{Spin ladder system with $l$ legs described by
  Hamiltonian ${\bf H}(h_{3},\tau_{l})$ [Eq.~(\ref{eq:8})]. }
\label{fig:tau_ladder}
\end{figure}

\begin{figure}[ht]
\centerline{\epsfig{file=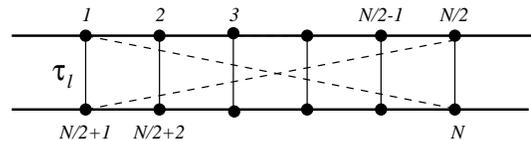,width=7cm}}
\caption{Spin ladder system with two legs described by
  Hamiltonian ${\bf H}(h_{3},\tau_{2})$.}
\label{fig:tau_2_ladder}
\end{figure}

Finally, let us mention that there is a second way to map ladder systems with $l$
legs onto one-dimensional systems with far reaching couplings:
\begin{equation}
  \label{eq:8}
  {\bf H}(h_{3},\tau_{l}) \equiv {\bf H}(h_{3}) + \tau_{l} {\bf H}_{N/l}.
\end{equation}
The couplings for the $l$ leg system are shown in Fig.~\ref{fig:tau_ladder},
%%%%%%%%%%%%%%%%%%%%%%%%%%%%%%%BEGIN-FIGURE%%%%%%%%%%%%%%%%%%%%%%%%%%%%%%%%
%%%%%%%%%%%%%%%%%%%%%%%%%%%%%%%%END-FIGURE%%%%%%%%%%%%%%%%%%%%%%%%%%%%%%%%%
and for the two leg system in Fig.~\ref{fig:tau_2_ladder},
%%%%%%%%%%%%%%%%%%%%%%%%%%%%%%%BEGIN-FIGURE%%%%%%%%%%%%%%%%%%%%%%%%%%%%%%%%
%%%%%%%%%%%%%%%%%%%%%%%%%%%%%%%%END-FIGURE%%%%%%%%%%%%%%%%%%%%%%%%%%%%%%%%%
The latter differs from the conventional two leg system (with periodic
boundary-conditions) only by a twist at the boundary, which should not change
the physical properties in the thermodynamical limit. Therefore, we expect a gap
in this system. The appearance of a gap in the systems with an even number $l$
of legs originates from the second term in the Hamiltonian~(\ref{eq:8}). If we
repeat the calculation of the expectation values~(\ref{eq:i3}) for the
Hamiltonian~(\ref{eq:8}) we find: 
\begin{eqnarray}
  \label{eq:9}
  \langle k | {\bf H} | k \rangle - \langle 0 | {\bf H} | 0 \rangle &=& 
  O(N^{-1}) \nonumber\\ && \hspace{-3cm} +
 \sum_{n=1}^{N} \left(
   \langle 0 | {\bf U}^{k} {\bf S}_{n}{\bf S}_{n+N/l} {\bf U}^{\dag^{k}}
              - {\bf S}_{n}{\bf S}_{n+N/l} | 0 \rangle 
 \right) = \nonumber \\ && \hspace{-3cm}
 \sum_{n=1}^{N} f_{l}^{k}\langle 0 | {\bf S}_{n}^{+}{\bf S}^{-}_{n+N/l} + 
                      {\bf S}_{n}^{-}{\bf S}^{+}_{n+N/l} | 0 \rangle+O(N^{-1}). 
\end{eqnarray}
The coefficient $f_{l}^{k}=[\cos(2\pi k/l)-1]\tau_{l}$ does not vanish unless 
\begin{equation}
  \label{eq:10}
  k=l,2l,\ldots.
\end{equation}
This means in particular that the LSM-operators ${\bf U}^{k}$, $k=1,\ldots,l-1$
do not create states, which are degenerate with the ground state in the
thermodynamical limit.

Those situations, where the ground state degeneracy is lifted completely, are of
special interest. They occur if the momenta of the states $|l\rangle={\bf
  U}^{l}|0\rangle$ [$p_{l}=l\pi(1-2M)+p_{s}$] and of the ground state
$|0\rangle$ ($p_{s}$) differ by a multiple of $2\pi$, i.e. for
\begin{equation}
  \label{eq:12}
  \frac{l}{2}(1-2M)\in \Bbb{Z}.
\end{equation}
The condition~(\ref{eq:12}) is satisfied exactly for the $l$ and $M$ values,
listed in Table~\ref{tab:legs}, i.e. for those values, where we expect a plateau
in the magnetization curve.
%%%%%%%%%%%%%%%%%%%%%%%%%%%%%%%%%%%%%%%%%%%%%%%%%%%%%%%%%%%%%%%%%%%%%%%%%%%%%%%%
%
\section{Conclusion and Perspectives}\label{sec:conclusion}
%
%%%%%%%%%%%%%%%%%%%%%%%%%%%%%%%%%%%%%%%%%%%%%%%%%%%%%%%%%%%%%%%%%%%%%%%%%%%%%%%%
In this paper we tried to elucidate the mechanism which generates gaps and
plateaus in spin $1/2$ antiferromagnetic Heisenberg models with nearest and next
to nearest neighbor couplings of strength $\alpha$.  A priori these models have
no gap in the presence of a homogeneous field $h_{3} > h_{3}^{c}(\alpha)$ above
the critical field $ h_{3}^{c}(\alpha)$, which is needed to surmount the singlet
triplet gap for $\alpha > \alpha_c = 0.241\ldots$.  The LSM construction
predicts the existence of soft modes (zero energy excitations) at wave vectors
$q^{(k)}(M)=k\pi(1-2M),\; k=1,2,3,\ldots$.

It is shown in Sec.~\ref{sec:LSM-construction} that the total spin squared ${\bf
  S}^2_T=S(S+1)$ of the lowest excited states at the soft mode momenta is the
same as that of the ground state for $S=M N$.  The soft modes are therefore
expected to generate signatures in the dimer dimer structure factor, since the
dimer operator does not change the total spin squared.

Indeed, for $M=1/4$ a pronounced peak is seen at the first soft mode
$q=q^{(1)}(1/4)=\pi/2$ if $\alpha \le 1/4$, indicating a large transition matrix
element $\langle 1|\bar {\bf D}(\pi/2)|0\rangle $ between the ground state -- with
momentum $p_{s}$ -- and the first excited state with momentum
$p_{s}+\pi/2$. 

The second soft mode $q=q^{(2)}(1/4)=\pi$, however, does not produce any visible
structure in the dimer dimer structure factor (for $\alpha \le 0.25$).  Here the
relevant transition matrix elements $\langle 1|{\bf D}(\pi)|0\rangle $ between
the ground state $|0\rangle$ and the first excited state with momentum
$p_{s}+\pi$ are small.  The situation is different for large
next-to-nearest-neighbor couplings $\alpha$.

The magnitude of the transition matrix elements $\langle 1|\bar {\bf
  D}(q)|0\rangle$ is crucial for the efficiency of the mechanism to generate a
plateau in the magnetization curve at a rational value of $M$ by means of a
periodic perturbation $\delta_{q}{\bf \bar D}(q)$.  According to the criterium
of Oshikawa, Yamanaka and Affleck\cite{OYA97} a plateau at $M$ is possible if
$q$ meets one of the soft modes $q=q^{(k)}(M),\; k=1,2,3,\ldots$.
  
Our numerical analysis shows that the width of the plateau -- i.e. the
difference of the upper and lower critical field (\ref{eq:47}) -- depends on the
magnitude of the transition matrix element in the unperturbed model
$(\delta_{q}=0)$. Indeed these matrix elements enter as initial conditions in
the differential equations [(2.4),(2.5)] and [(2.2),(2.3)] in
Refs.~\onlinecite{FKM98a,FKM98b}, which describe the evolution of gaps and
plateaus under the influence of a periodic perturbation $\delta_{q}{\bf \bar
  D}(q)$. We have also demonstrated that a superposition of two periodic
perturbations [$\bar {\bf D}(2\pi/3) +\bar {\bf D}(\pi/3) $] generates two
plateaus in the magnetization curve exactly at those magnetization values
($M=1/6,1/3$) where the period ($q=2\pi/3,\pi/3$) in the perturbation coincides
with the first soft mode momentum $q=q_{3}(M)$.

Ladder systems with $l$ legs [cf. Fig.~\ref{fig:l_ladder}] can be interpreted as
one-dimensional systems with additional couplings over $l$ lattice spacings and
a dimerized perturbation~(\ref{eq:dimer_l}) and~(\ref{eq:2}). The latter breaks
translation invariance of the 1D system and the Fourier analysis~(\ref{eq:7}) of
the dimerized perturbation~(\ref{eq:dimer_l}) reveals the occurrence of
magnetization plateaus [cf. Tab.~\ref{tab:legs}] in spin ladder systems with $l$
legs.

In this paper we concentrated our investigations on different spin-1/2 systems
that all had in common that they reduce in the unperturbed case to critical
Heisenberg chains. It should be pointed out that there exist exact solutions on
other (multichain spin-1/2 antiferromagnetic Heisenberg-like) models,\cite{PZ93}
where the existence of magnetization plateaus is still unclear. We remark that
the application of the method we presented here, is not limited to the cases we
discussed. However, besides looking for the existence of soft modes it turns out
to be needed to discuss the strength of the transition matrix elements. Here, we
cannot offer general prescription and therefore the dynamics of each different
model of interest has to be treated separately.
%%%%%%%%%%%%%%%%%%%%%%%%%%%%%%%%%%%%%%%%%%%%%%%%%%%%%%%%%%%%%%%%%%%%%%%%%%%%%%%%
% 
\acknowledgments
%
%%%%%%%%%%%%%%%%%%%%%%%%%%%%%%%%%%%%%%%%%%%%%%%%%%%%%%%%%%%%%%%%%%%%%%%%%%%%%%%%
We would like to thank A. Honecker and A. Kl\"umper for discussions.
%%%%%%%%%%%%%%%%%%%%%%%%%%%%%%%%%%%%%%%%%%%%%%%%%%%%%%%%%%%%%%%%%%%%%%%%%%%%%%%%
% 
%\section*{References}

%\bibliography{/home/karbach/REFERENCES/references}
%\bibliographystyle{prsty}
%\begin{thebibliography}{99}
%
%%%%%%%%%%%%%%%%%%%%%%%%%%%%%%%%%%%%%%%%%%%%%%%%%%%%%%%%%%%%%%%%%%%%%%%%%%%%%%%%
\vspace{12cm}
%%%%%%%%%%%%%%%%%%%%%%%%%%%%%%%%%%%%TABLE%%%%%%%%%%%%%%%%%%%%%%%%%%%%%%%%%%%%%
\begin{table}[ht]
  \begin{tabular}{c|cccccc}
    $l$&2&3&4&5&6 \\ \hline
    $M$&0&1/6&0;1/4&1/10;3/10&0;1/6;1/3
  \end{tabular}
  \caption{The possible position of plateaus for spin ladders~(\ref{eq:H_h3_al})
    with $l$ legs.}
  \label{tab:legs}
\end{table}
%%%%%%%%%%%%%%%%%%%%%%%%%%%%%%%%%%%%TABLE%%%%%%%%%%%%%%%%%%%%%%%%%%%%%%%%%%%%%

\end{document}